\documentclass[oneside,twocolumn]{article}

\usepackage{graphics}
\usepackage{graphicx}

\begin{document}
\title{Collective dynamics of evanescently coupled excitable lasers with saturable absorber}
\author{\small{M. Lamperti$^1$ and A. M. Perego$^2$} \\
\small{$^{1,*}$ Politecnico di Milano, Department of Physics  and IFN-CNR, Via G. Previati 1/C, 23900, Lecco, Italy} \\
\small{$^2$ Aston Institute of Photonic Technologies, Aston University, Birmingham B4 7ET, Aston Street, UK} \\
\small{$^*$\texttt{marco1.lamperti@polimi.it}}
}
\date{}
\maketitle
\abstract{
We present a numerical study of the collective dynamics in a population of coupled excitable lasers with saturable absorber. At variance with previous studies where real-valued (lossy) coupling was considered, we focus here on the purely imaginary coupling (evanescent wave coupling).
We show that evanescently coupled excitable lasers synchronize in a more efficient way compared to the lossy coupled ones. Furthermore we show that out-of-diagonal disorder-induced localization of excitability takes place for imaginary coupling too, but it can be frustrated by nonvanishing linewidth enhancement factor.
} 

\section{Introduction}
\label{intro}

The last decade has been characterized by the rise and spread of data storage and analytics over many industrial and commercial fields, due to the decreasing cost of data collection, storage and transmission.
While it is clear that large datasets are needed to obtain meaningful statistical information about complex systems, the instruments used for such information distillation have not evolved with the same speed as the data infrastructure, due on one hand to our still rough understanding of large, complex systems, and on the other hand to the exponentially increasing computing resources needed to apply more and more complex models for data interpretation.
A promising framework for advanced data processing tasks is constituted by neural networks \cite{furber2016large}, which currently see widespread usage to perform actions that are traditionally a premise of human beings, such as pattern recognition, natural language processing (information extraction from texts and translation), and identification of objects in images.
Although specialized hardware and software has been developed to increment the efficiency of implementing such framework with the currently most advanced computing paradigma, based on a Von Neumann architecture executed on CMOS-based integrated circuits, the level of complexity and efficiency of the human brain is far from reach by many orders of magnitude.

The maturity and apparent limits of the current technology have pushed scientists from many disciplines to propose new computing paradigms that could enable a leap towards the higher processing power required for the aforementioned purposes; the most promising results take inspiration from what is considered to be the most advanced naturally evolved computer: the human brain.
The resulting field of neuromorphic computing aims at taking advantage from systems that exhibit naturally the characteristics that make biological neural networks so powerful and efficient, mapping the process paradigm to its underlying dynamics rather than abstracting away from it, as it is currently done with software implementation of neural networks on serialized, digital hardware \cite{shastri2017principles}.

The two key ingredients for implementing neuromorphic computing are neuron-like behavior and a large network of interconnections \cite{furber2016large}.
The first ingredient is provided by excitable systems: excitability is an ubiquitous process in nature mostly known in biology \cite{Murray,Izhikevich} in the context of the neuronal cell activity and can be defined as the generation of spike-like behaviour in one or more system dynamical variables in response to external perturbations whose magnitude exceeds a given threshold.
In such regime the system response does not depend on the perturbation strength and each spike generation is followed by a refractory time during which the system remains silent; after such refractory time the emission of another spike can take place again.
Excitability needs to be provided by a suitable neuron-like system, which in turn will constitute the building block to perform simple computations in a highly parallel fashion, making the system fast and efficient. What is needed to perform complex computations is a large network of interconnects that weight the outputs of previous neurons and route them to other computation elements, i.e. other neuron-like blocks.
A network of interconnects is thus the key ingredient to transform a bunch of excitable systems into a highly-efficient computer able to solve complex tasks.

Currently, the golden standard for electronics-based computing, CMOS integrated circuits, is unable to provide these two elements while keeping the power efficiency high \cite{schuller+stevens}.
As we approach the few-atom transistor, with bigger and bigger challenges without any significant decrease of power consumption, new directions have been probed in search for another platform that could provide high integration, low power consumption and scalability.
Thanks to recent advances in optoelectronics integration, photonics technologies constitute one of the most promising platforms for neuromorphic computing: neuromorphic photonics is gaining momentum as a research field where emerging photonics technologies are used to mimic neuronal dynamics and/or to perform computational tasks based on brain inspired strategies \cite{PrucnalBook}.
Excitability in photonics has been reported in a variety of different lasers and amplifiers systems \cite{Lenstra,Barbay,Barland,Giudici,Arecchi,Brunstein,TurconiPRL,ShastriSRep,Romeira}.
Possibly the first historically studied example of a photonics system exhibiting excitability is the semiconductor laser with saturable absorber \cite{Lenstra}.
Excitability in such laser is enabled by a phase space portrait exhibiting a limit cycle close to a saddle node bifurcation.
In the regime where the absorption is the slow dynamical variable, if the below, but close-to-threshold laser is perturbed strongly enough, then the stimulated emission process builds up producing the emission of giant light pulse.
Such emission depletes the gain and is followed by a refractory time after which the laser is ready to be excited again. 
Such dynamics can be associated to a so called type III excitability \cite{Barbay}.

Engineering optically excitable elements connectivities and studying their collective properties and dynamics are both crucial tasks towards achieving a general understanding of neuromorphic photonics systems.
Focussing our attention to the excitable laser with saturable absorber, it is important to stress that coupling engineering of excitable semiconductor lasers with saturable absorber has been shown to allow pattern recognition \cite{ShastriSRep},  neuronal circuits design \cite{Shastri2} and coincidence detection devices \cite{Shastri}.
As far as the collective dynamics is concerned, we have shown theoretically in two recent works \cite{Perego,Lamperti} that temporal and intensity synchronization, array enhanced coherence resonance, and even disorder-induced localization of excitability can take place in arrays of excitable lasers with nearest neighbour real-valued (lossy) coupling.
In this work we extend our previous studies on synchronization and disorder-induced localization of excitability to the case of an ensemble of excitable lasers with saturable absorber coupled via a nearest neighbour purely imaginary coupling coefficient. Imaginary valued couplings describe the physical evanescent waves interaction between adjacent laser cavities. Such interaction  is most likely the relevant coupling mechanism for micropillars lasers, where collective excitable dynamics could be observed \cite{Selmi}.

\section{The model}
\label{sec:model}
We consider here a population of $n$ lasers with nearest neighbour coupling. The following normalized Yamada model describes the $i$-th laser dynamics:\newline
\begin{eqnarray}
\label{eq1} \nonumber 
\dot{F}_i&=&\frac{1}{2}[G_i(1-i\alpha)-Q_i(1-i\beta)-1]F_i+\sigma_i\\ \nonumber&-&i(K_{i,i+1}+K_{i,i-1})F_{i}+iK_{i+1,i}F_{i+1}+iK_{i-1,i}F_{i-1},\\ \nonumber
\dot{G}_i&=&\gamma(A-G_i-I_iG_i),\\ 
\dot{Q}_i&=&\gamma(B-Q_i-aQ_i I_i ).
\end{eqnarray}

\ \newline
Here $F_i$ denotes the electric field strength, $I_i=|F_i|^2$ its intensity, $G_i$  and $Q_i$ gain and absorption respectively. $A$ is the pump parameter, $B$ the background absorption, $a$ the differential absorption relative to the differential gain, $\gamma$ is the absorber and gain decay rate, $\alpha$ and $\beta$ denote the linewidth enhancement factors for the gain and the absorber respectively; these parameters do not have subscript $i$ since they have been considered identical for all lasers.
$\sigma_i$ describes a delta correlated Gaussian noise term of strength $D$ with $\langle \sigma_i(t_1)\sigma_j(t_2) \rangle=\sqrt{2D}\delta(t_1-t_2)\delta_{ij}$ providing the perturbations needed for excitable behaviour. The dot denotes temporal derivative and the time variable has been normalized to the uncoupled laser photon lifetime. $K_{i,j}$ denotes the nearest neighbour coupling strength describing light coupling from the $i$-th to the $j$-th laser where the reciprocity condition $K_{i,j}=K_{j,i}$ has been imposed. The identical coupling case $K_{i,i+1}=K$ results in an effective discrete Laplace diffraction operator in the array. \newline
Periodic conditions at the array boundary have been applied. Across the whole paper we have set $A=6.5$, $B=5.8$, $a=1.8$, $\gamma=10^{-3}$ while for $\alpha$ and $\beta$ different values have been used  and specified across the paper.

\section{Synchronization}
\label{sec:synchronization}
\begin{figure}
  \includegraphics[width=\columnwidth]{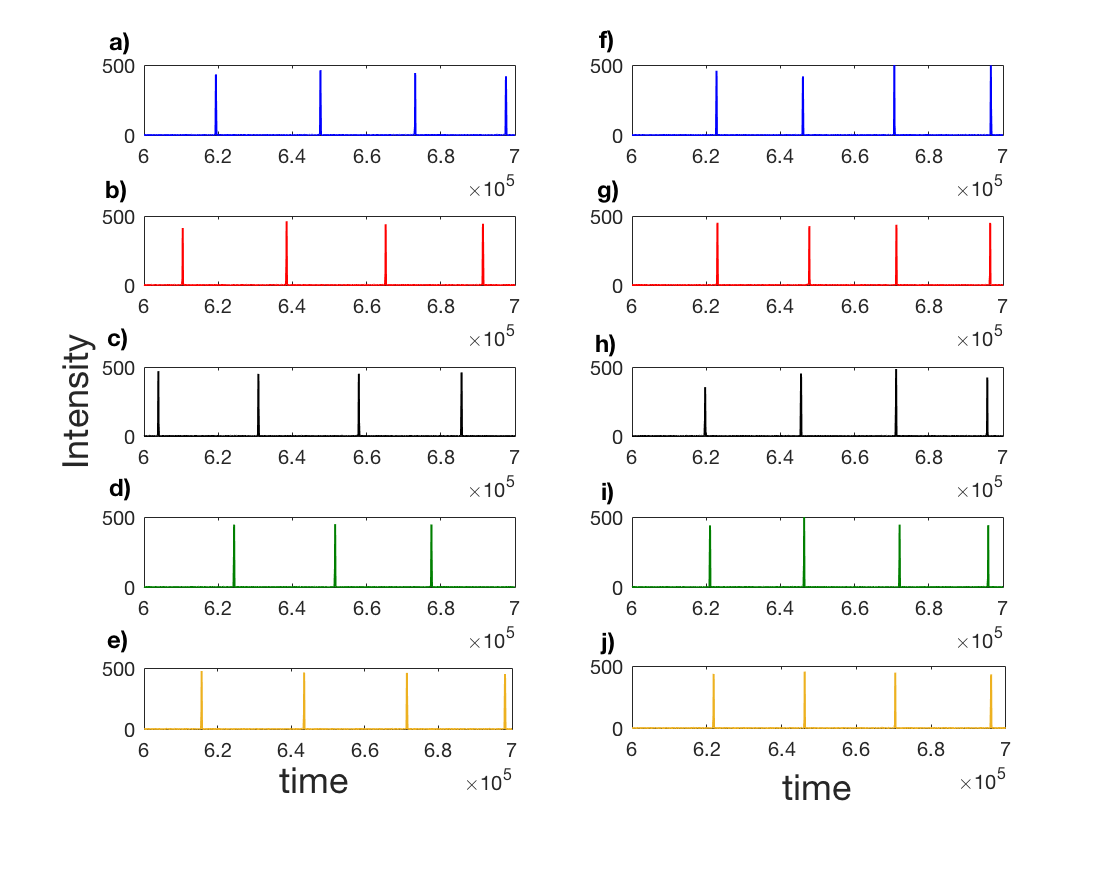}
\caption{In panels a) to e) the temporal traces of five different uncoupled lasers have been plotted showing uncorrelated lasers firing. In panels f) to j) the time traces of five randomly-picked up lasers in a chain of 50 coupled elements have been depicted. Spatial synchronization is clearly evident in presence of non vanishing coupling. Parameters used are $D=0.01$, $\alpha=\beta=0$ for both the coupled and uncoupled case; $K_0=0$ in a) to e) and $K=0.05$ in f) to j).}
\label{fig:1}
\end{figure}

We have first studied the effect of coupling on the synchronization of the firing events in the lasers array as a function of the number of lasers constituting the array itself and of the values of the linewidth enhancement factors. To this purpose we have considered independent additive noise sources to be present in all lasers and varied their amplitude, $D$, from 0 to 0.15. The coupling strength, $K=K_0$, has been considered identical for all lasers and has been varied from 0 to 1. For each pair $(D,K)$ we run one simulation for a time duration $T=100000$.
The pulse synchronization can be understood both in space and time: in the first case it refers to neighbor lasers emitting a pulse at the same time while in the second case it characterizes a regular (i.e. equally spaced) emission of pulses from the single laser. To give a qualitative flavour of temporally synchronous pulses (spikes) emission of coupled lasers we have plotted in Fig.~\ref{fig:1} the temporal traces corresponding to coupled and uncoupled laser (left and right column, respectively).

We have characterized the spatial synchronization by using an indicator ($S$) that quantifies the phase slip between the firing events of nearest-neighbors lasers \cite{Perego,Rosenblum}.
The phase of the $i$-th laser reads:
\begin{eqnarray}
\phi_i(t)=\frac{t-\tau_k}{\tau_{k+1}-\tau_{k}}+2k\pi
\end{eqnarray}
where $\tau_k$ is the time of the $k$-th firing event, i.e. the position in time of the $k$-th emitted pulse. We define then
\begin{eqnarray}
s_i=\sin\left(\frac{\phi_i-\phi_{i+1}}{2}\right)^2
\end{eqnarray}
which characterizes the phase synchronization between the $i$-th and $i+1$-th lasers. The average both across all the array elements and along temporal duration of the time trace gives the $S$ indicator, which provides a measure of the spatial synchronization of the whole system:
\begin{eqnarray}
S=\lim_{T \to \infty}\frac{1}{T}\int_0^T\left(\frac{1}{n}\sum_{i=1}^ns_i\right)dt.
\end{eqnarray}
The maximum synchronization occurs for $S=0$ while for the completely non synchronized state $S=0.5$.

We have furthermore characterized the regular emission of pulses of the individual laser (temporal synchronization) by means of the single laser pulse jitter defined as $J=\sigma_T/\langle T \rangle$, where $\langle T \rangle$ is the average time interval between two consecutive pulses and $\sigma_T$ is the standard deviation. A value of $J$ close to 0 indicates regular pulses emission, while values close or greater than 1 indicate poor regularity.
\begin{figure}
  \includegraphics[width=\columnwidth]{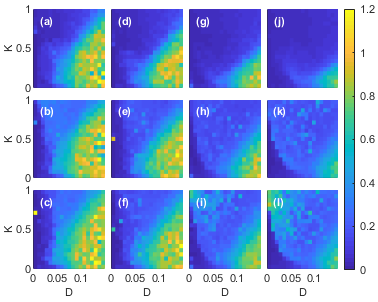}
\caption{The single-laser temporal synchronization index, $J$, is plotted in the $(D,K)$ space for different numbers of lasers (rows - 4, 20 and 150 from top to bottom) and different values of the line enhancement factors (columns - 0, 1, 3 and 5 from left to right).}
\label{fig:2}
\end{figure}

\begin{figure}
  \includegraphics[width=\columnwidth]{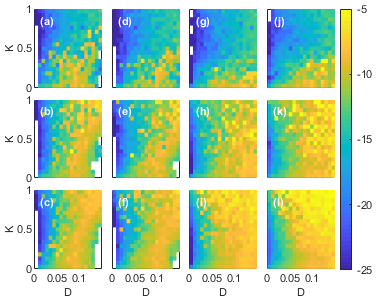}
\caption{The logarithm of the spatial synchronization, $\log_{10}(S)$, is plotted in the $(D,K)$ space for different numbers of lasers (rows - 4, 20 and 150 from top to bottom) and different values of the line enhancement factors (columns - 0, 1, 3 and 5 from left to right).}
\label{fig:3}
\end{figure}

The panels in Fig.~\ref{fig:2} and in Fig.~\ref{fig:3} show the $J$ and $S$ synchronization indicators, respectively, for different number of lasers constituting the chain and different values of the line enhancement factors, versus the noise and the coupling strength.
The behavior observed here is qualitatively different from the case of real (dissipative) coupling, showing no array-enhanced coherence resonance \cite{Perego}.
Instead of enlarging the region of the $(D,K)$ parameter space characterized by high synchronization, an increase in the number of coupled lasers improves the temporal synchronization in a limited region of the parameters space (the lower-left one) while worsening the synchronization in all the remaining region. In other words, enlarging the array creates a sharper separation between the regions of high and low synchronization, without changing its area. \newline
On the other hand, increasing the number of lasers in the array has the effect of reducing the region of highest spatial synchronization, i.e. low value of the $S$ indicator. This effect is well visible in the part of the $(D,K)$ space characterized by $D > 0.05$, and manifests itself quickly with the increase of lasers forming the chain (compare first and second row of Fig.~\ref{fig:3}). It is worth noting that even in the worst synchronization cases, the $S$ indicator is smaller than $10^{-5}$, indicating good spatial synchronization in all the $(D,K)$ space.

The second control parameter we considered, the line enhancement factor, differently from the number of lasers modifies the shape and size of the synchronization region. For a small number of lasers (4, first row in Figs. \ref{fig:2}-\ref{fig:3}), its progressive increase from 0 to 5 causes an enlargement of the region of high synchronization towards larger values of the noise, both in time and space. For a longer chain (20 and 150 lasers, two bottom rows), high synchronization occurs in a more and more localized region towards small values of $K$ and $D$; even in a strong coupling regime high temporal synchronization of spikes emitted by the same laser results more difficult to achieve. 

As compared to the real coupling case \cite{Perego} we can observe that a much smaller value of the coupling strength is needed to achieve comparable synchronization. Imaginary coupling is hence a much more efficient way to achieve synchronization of excitability.

\section{Disorder-induced localization}
In our previous study for real-valued couplings we pointed out how randomness in the laser array coupling strength is strikingly able to induce spatial exponential localization of excitable behavior in a given area of the array \cite{Lamperti}. Such disorder-induced localization does not occur due to a trivial breaking of the lasers chain but due to a dynamical process entailing both scattering and dissipation. We want to stress that the localization we are discussing here, although mediated by disorder, can not be assimilated directly to the celebrated \emph{Anderson localization} \cite{Anderson}. The latter, although it has inspired our work, is occurring in ideally conservative systems while in our system dissipation plays of course a paramount important role. Still inspired by the conspicuous literature on \emph{Anderson localization} we can borrow some useful terminology. In the traditional studies of \emph{Anderson localization} of electronic transport in solid state physics, the distinction between diagonal and out-of-diagonal disorder is made when referring to disorder on the individual atomic sites, or in the hopping terms between neighbour atomic sites respectively \cite{Pendry}. In our system we identify clearly the former as corresponding to randomness in some internal laser parameter (e.g. pump parameter), while the latter to the coupling between the lasers. In this paper we will consider out-of-diagonal disorder defining the coupling as:
$K_{i,i+1} = K_0 + \rho_{i,i+1}$, being $K_0$ a value common to all lasers and $\rho_{i,i+1}$ a random number constant in time drawn from a uniform distribution in the interval $[-r,+r]$ which describes light coupling from the $i$-th laser to nearest neighbour on the right. We recall also that the reciprocity constraint implies $\rho_{i,i+1}=\rho_{i+1,i}$.
\begin{figure}
  \includegraphics[width=\columnwidth]{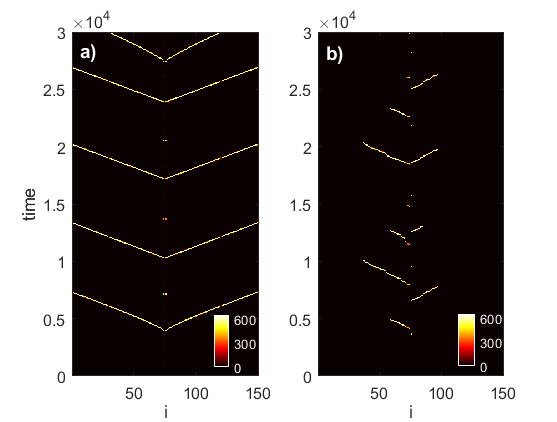}
\caption{In a) the intensity dynamics in absence disorder: an excitability wave propagates through the array. In b) an example of the temporal dynamics in presence of disorder is shown: excitability is spatially localized. Parameters used are: $\alpha=\beta=0$, $K_0=0.05$, $r=0.005$.}
\label{fig:4}
\end{figure}
\newline In order to investigate the disorder-induced localization we have considered that the additive noise is present, without loss of generality, only in the central laser of the array. If all the lasers are coupled with identical coupling strength ($K_{i,i+1}=K_0$ $\forall$ $i$), then an excitability wave propagates from the centre of the array both towards left and right (see Fig.~\ref{fig:4}).
We have considered a chain of 150 lasers with an average coupling $K_0 = 0.05$, a noise on the central laser with strength $D=0.1$ and a random coupling distribution varying in the interval $r \in [0, 0.002]$. We have verified that in this condition, with noise acting only on one laser, we have regular emission of pulses that propagate to all the lasers in the chain for $r = 0$.
\begin{figure}
\includegraphics[width=\columnwidth]{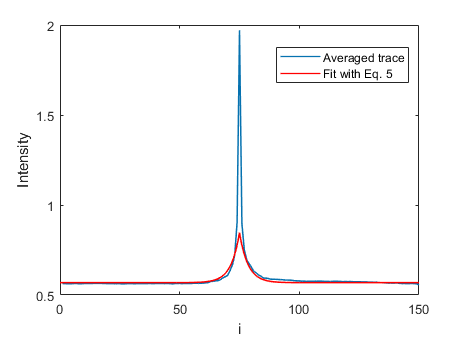}
\caption{The temporal average of 100 traces characterized by different realizations of disorder with the same value of $r$ is plotted here together with the fit to Eq.~\ref{eq:fit} used to extract the value of the localization exponent, $\lambda$.
To extract the correct localization exponent we discard the central laser, whose average intensity is systematically too high due to the presence of additive noise, and its nearest neighbors.
Parameters used are the same of Fig.~\ref{fig:5}, with $\alpha=\beta=0$ and $r=0.002$.
}
\label{fig:fit}
\end{figure}

\begin{figure}
\includegraphics[width=\columnwidth]{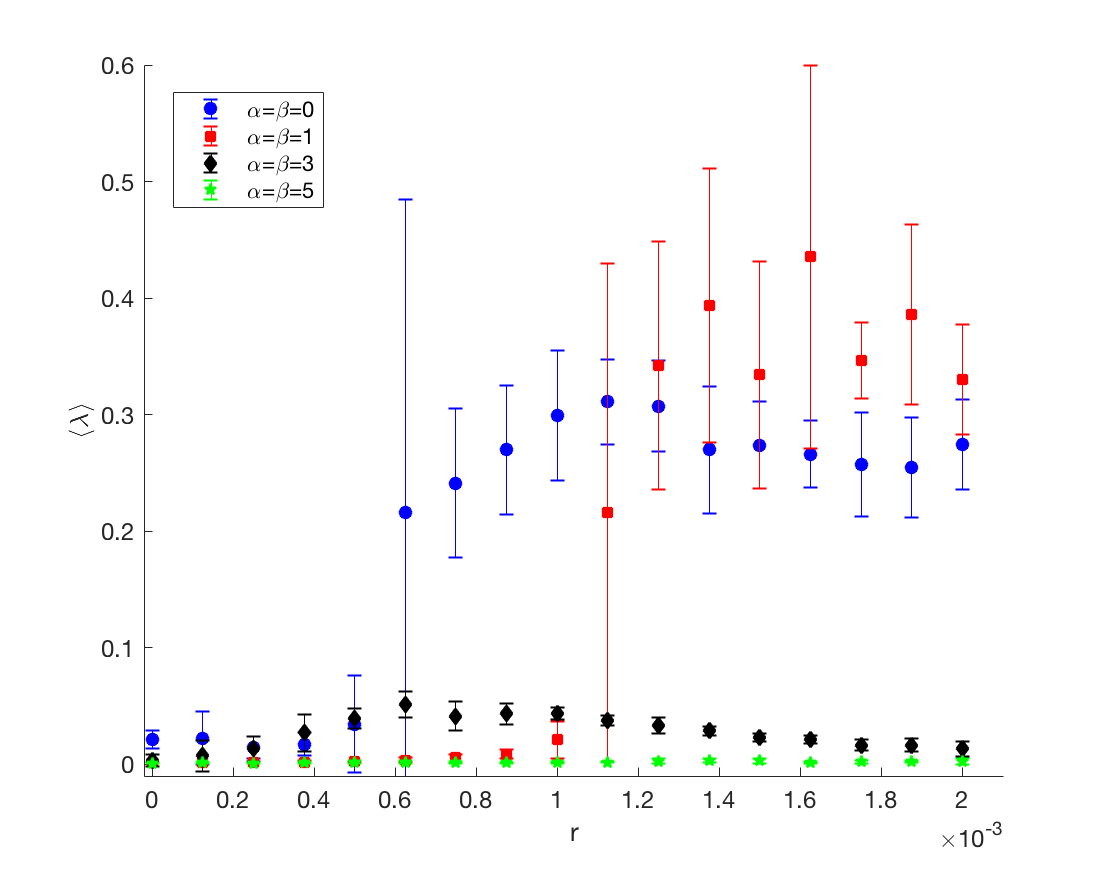}
\caption{The average localization exponent $\langle\lambda\rangle$ has been plotted versus randomness strength $r$ for 3 different values of the linewidth enhancement factors $\alpha$ and $\beta$. For vanishing $\alpha$ and $\beta$ a transition from ballistic to localized regime occurs, low values of $\alpha$ and $\beta$ require an higher threshold for localization. At large values of the linewidth enhancement factor localization is lost. Parameters used are $D=0.1$ and $K=0.05$.}
\label{fig:5}
\end{figure}

For each value of $r$, 100 numerical experiments with a time duration of $T=50000$ has been carried out. For each disordered coupling configuration we have computed the average intensity trace across the array. We have then fitted the averaged intensity distribution after the 100 realizations of the disorder with the same strength $r$, with the following exponential function:

\begin{eqnarray}
f=b+\exp\left(-\lambda |i-i_0|\right)
\label{eq:fit}
\end{eqnarray}
where $i_0$ denotes the position of the central laser (see Fig.~\ref{fig:fit} for an example).
This procedure has been repeated 8 times, so that the average value and standard deviation of the fit parameters can be extracted; the average localization exponent $\langle\lambda\rangle$ versus the disorder strength $r$ is plotted in Fig.~\ref{fig:5}, together with error bars indicating the standard deviation. 

Such disorder-induced localization occurs for coupling values such that if we couple all the lasers with identical value of $K_0-r$ equal to the minimum possible value generated by the random process, we would still have an excitability wave spreading from the array center both towards left and right; this ensures the fact that localization is not caused by a trivial breaking of the laser chain but is indeed provoked by a dynamical effect.
For vanishing $\alpha$ and $\beta$ we can notice that localization occurs for coupling strength much smaller than the ones needed in the real coupling case (about 0.3, see \cite{Lamperti}).\newline
We have furthermore investigated the impact of linewidth enhancement factor both for the gain medium ($\alpha$) and for the saturable absorber ($\beta$) on the disorder-induced localization of excitability. Physically $\alpha$ and $\beta$ are responsible for coupling of amplitude to phase fluctuations.
It is interesting to appreciate that linewidth enhancement factor ($\alpha$,$\beta\ne 0$) causes a decrease (for $\alpha=\beta=1$) and even a vanishing (for $\alpha=\beta=3$ or $5$) of the localization.

\section{Conclusions}
We have demonstrated with the help of numerical simulations that evanescently coupled excitable lasers with saturable absorber synchronize in a more efficient way compared to lossy coupled ones, but that they do not exhibit array enhanced coherence resonance. We have furthermore demonstrated that disorder-induced localization of excitability caused by randomness in the coupling strength exists for imaginary coupling too. Taking into account the linewidth enhancement factor both in the gain medium and in the saturable absorber we have shown that localization is reduced and eventually quenched by an increase of the linewidth enhancement factor. This fact may constitute a serious obstacle in the experimental observation of the disorder-induced localization of excitability.
Our results shed further light on the collective dynamics of coupled excitable lasers with saturable absorber and suggest interesting directions for experimental studies.

\section{Acknowledgements}
The authors gratefully acknowledge useful discussion with Dr. Sylvain Barabay.

\bibliographystyle{}
\bibliography{biblio}

\begin{thebibliography}{23}

\bibitem{furber2016large}
S.~Furber, Journal of neural engineering \textbf{13}, 051001 (2016)

\bibitem{shastri2017principles}
B.J. Shastri, A.N. Tait, T.F. de~Lima, M.A. Nahmias, H.T. Peng, P.R. Prucnal,
  \emph{Principles of neuromorphic photonics} (2017), \texttt{1801.00016}

\bibitem{Murray}
J.D. Murray, \emph{Mathematical Biology: I. An Introduction}, Vol.~1
  (Springer-Verlag, 3rd Edition Berlin Heidelberg, 2002)

\bibitem{Izhikevich}
E.M. Izhikevich, International Journal of Bifurcation and Chaos \textbf{10},
  1171 (2000)

\bibitem{schuller+stevens}
I.K. Schuller, R.~Stevens, Tech. rep. (2015),
  \texttt{https://science.energy.gov/$\sim$/media/bes/
  pdf/reports/2016/NCFMtSA\_rpt.pdf}

\bibitem{PrucnalBook}
P.R. Prucnal, B.V. Shastri, \emph{Neuromorphic Photonics} (Taylor \& Francis,
  1st Edition Boca Raton, 2017)

\bibitem{Lenstra}
J.L. Dubbeldam, B.~Krauskopf, D.~Lenstra, Physical Review E \textbf{60}, 6580
  (1999)

\bibitem{Barbay}
S.~Barbay, R.~Kuszelewicz, A.M. Yacomotti, Optics letters \textbf{36}, 4476
  (2011)

\bibitem{Barland}
S.~Barland, O.~Piro, M.~Giudici, J.R. Tredicce, S.~Balle, Physical Review E
  \textbf{68}, 036209 (2003)

\bibitem{Giudici}
M.~Giudici, C.~Green, G.~Giacomelli, U.~Nespolo, J.~Tredicce, Physical Review E
  \textbf{55}, 6414 (1997)

\bibitem{Arecchi}
F.~Plaza, M.~Velarde, F.~Arecchi, S.~Boccaletti, M.~Ciofini, R.~Meucci, EPL
  (Europhysics Letters) \textbf{38}, 85 (1997)

\bibitem{Brunstein}
M.~Brunstein, A.M. Yacomotti, I.~Sagnes, F.~Raineri, L.~Bigot, A.~Levenson,
  Physical Review A \textbf{85}, 031803 (2012)

\bibitem{TurconiPRL}
M.~Turconi, M.~Giudici, S.~Barland, Physical Review Letters \textbf{111},
  233901 (2013)

\bibitem{ShastriSRep}
B.J. Shastri, M.A. Nahmias, A.N. Tait, A.W. Rodriguez, B.~Wu, P.R. Prucnal,
  Scientific Reports \textbf{6}, 19126 (2016)

\bibitem{Romeira}
B.~Romeira, R.~Av{\'o}, J.M. Figueiredo, S.~Barland, J.~Javaloyes, Scientific
  reports \textbf{6}, 19510 (2016)

\bibitem{Shastri2}
B.J. Shastri, M.A. Nahmias, A.N. Tait, B.~Wu, P.R. Prucnal, Optics express
  \textbf{23}, 8029 (2015)

\bibitem{Shastri}
B.J. Shastri, A.N. Tait, M.~Nahmias, B.~Wu, P.~Prucnal, \emph{Coincidence
  detection with graphene excitable laser}, in \emph{CLEO: Science and
  Innovations} (Optical Society of America, 2014), pp. STu3I--5

\bibitem{Perego}
A.M. Perego, M.~Lamperti, Physical Review A \textbf{94}, 033839 (2016)

\bibitem{Lamperti}
M.~Lamperti, A.M. Perego, Physical Review A \textbf{96}, 041803(R) (2017)

\bibitem{Selmi}
F.~Selmi, R.~Braive, G.~Beaudoin, I.~Sagnes, R.~Kuszelewicz, S.~Barbay, Opt.
  Lett. \textbf{40}, 5690 (2015)

\bibitem{Rosenblum}
M.G. Rosenblum, A.S. Pikovsky, J.~Kurths, Circuits and Systems I: Fundamental
  Theory and Applications, IEEE Transactions on \textbf{44}, 874 (1997)

\bibitem{Anderson}
P.W. Anderson, Physical Review \textbf{109}, 1492 (1958)

\bibitem{Pendry}
J.B. Pendry, J. Phys. C: Solid State Phys. \textbf{15}, 5773 (1982)

\end{thebibliography}

\end{document}